\documentclass[11pt]{article}

\usepackage[margin=1in]{geometry}

\usepackage{graphics}
\usepackage{epsfig}
\usepackage{amsmath}
\usepackage{array}
\begin{document}
\begin{center}
{\LARGE Constraining the bulk viscous coefficients in a viscous universe with cosmological constant \\[0.2in]}
{ Athira Sasidharan$^*$ and Titus K Mathew$^+$\\
e-mail:athirasnair91@gmail.com, titus@cusat.ac.in \\ $^*$Department of
Physics, NSS Hindu College, Changanacherry, Kerala, India \\$^+$Department of
Physics, Cochin University of Science and Technology, India,.}
\end{center}

\abstract{In this paper we consider dissipative effects in $\Lambda$CDM model, i.e., we consider a universe with cosmological constant having viscous matter. We assume the most general form for bulk
viscous coefficient,
$\zeta=\zeta_{0}+\zeta_{1}\frac{\dot{a}}{a}+\zeta_{2}\frac{\ddot{a}}{\dot{a}}$ and obtained various constrains for $\zeta$'s . We also studied the background study of the model with $\zeta=\zeta_{0}$ and $\zeta=\zeta_{1}\frac{\dot{a}}{a}$. Extracted the value of $\zeta_1$ using the Pantheon data and also obtained its thermodynamic evolution and the age.}

\section{Introduction}
\label{sec:intro}
Since the discovery of accelerating universe \cite{Riess1,Perl1,Bennet1,Tegmark1,Seljak,Komatsu1}, active research is been taking place looking for the cause producing the acceleration and also for a model that would incorporate this acceleration. To the present, there are many models that fits this acceleration. Of these, the simplest and the most successful is the interpretation of dark energy as the cosmological constant. However, the discrepancy between the observed and calculated values of dark energy density, known as cosmological constant problem\cite{Weinberg,Carroll}, and unexplained coincidence of two dark sectors-dark energy and dark matter,  known as Cosmic Coincidence problem \cite{Zlatev}, make rooms for other models in explaining the current acceleration. Some of these models include quintessence
\cite{fujii,carroll}, k-essence \cite{chiba1} and perfect fluid
models (like Chaplygin gas model) \cite{kamen1}, $f(R)$ gravity \cite{capo1}, $f(T)$\
gravity \cite{ferraro1}, Gauss-Bonnet theory \cite{nojiri}, Lovelock
gravity \cite{pad2}, Horava-Lifshitz gravity \cite{horava1},
scalar-tensor theories \cite{amendola1}, braneworld models
\cite{dvali1} etc.

A less complicated unified dark energy model is the bulk viscous models. In \cite{fabris1,li1,Hiplito1,av1,av2,Athira1,Jerin1}, bulk viscous matter dominated universe is considered and was found that this viscosity alone can produce acceleration in the expansion of the universe. Phase space analysis of this model indicates that only viscosity with constant bulk viscous coefficient predicts all the conventional phases of the universe i.e., a prior radiation dominated phase, followed by a decelerated matter dominated phase and then finally evolving to a de Sitter type universe \cite{Athira2}. A bayesian analysis of this model shows that it is not so superior over the $\Lambda$CDM model, but have only a slight advantage over it\cite{Athira3}. However Maartens \cite{Maartens} has pointed out that these viscous models violates near equilibrium condition (NEC),
\begin{equation}
\label{NEC}
\frac{\Pi}{P}\ll 1
\end{equation}
There are works \cite{Riess1,Tegmark1} showing that $\Lambda$ is an inevitable content of the universe. The matter content of the universe has disspations so it is worth full to consider a universe filled wth viscous matter having a cosmological constant \cite{Gron}. Also recent papers \cite{Cruz, Cruz1} showed that introducing $\Lambda$ with viscosity can attain this NEC. We neglect the other dissipative phenomena like shear viscosity as it is inconsistent with the isotropic nature of the universe. So the only viscosity component to be considered is the bulk viscosity.

In this paper we first analyse the basic formalism  of the bulk viscous matter dominated universe with cosmological constant. We consider the general form for the bulk viscous coefficient and using Eckart formalism, obtain the expression for Hubble Parameter and the scale factor. We also analyse the equation of state parameter and the deceleration parameter and from the behavior of these parameters, constrains on the viscous parameters was obtained. In section \ref{sec 2}, we did the background study of the model for constant viscosity and constrains on the parameter is also obtained. We also analysed the age, thermodynamic behavior and asymptotic behavior of the model. In section \ref{sec 3},we consider the viscous coefficient as a function of Hubble parameter, i.e., $\zeta=\zeta_1 H$, extracted the value of $\zeta_1$ and studied the background evolution and cosmological parameters and the age of the universe. In section \ref{sec 4}, the results and conclusion are discussed.

\section{Viscous matter with cosmological constant}
We consider a spatially flat universe described by FLRW metric. We assume that the universe contains viscous  matter (both dark and baryonic) and cosmological constant as dark energy. We neglect the radiation component since its percentage composition is very small and also we are dealing with the late time acceleration. Eckart formalism \cite{Eckart1,weinberg2} is used for the bulk viscous pressure and is given by,
\begin{equation}
\label{p} P^{*}=P-3\zeta H
\end{equation}
where $P$ is the normal pressure, which we assume as zero for the whole matter component of the universe (both dark and baryonic) and $\zeta$ is the coefficient of bulk viscosity. So the effective pressure will only be that from the bulk viscosity. The coefficient $\zeta$ is basically a transport coefficient, hence it would depend on the dynamics of the cosmic fluid. Since the exact form of $\zeta$ is unknown, we consider the most general form for the bulk viscous coefficient $\zeta$     \cite{Athira1,Avelino,Athira2,ren1,Singh}, which is a linear combination of the three terms as,
\begin{equation}
\label{zeta}
\zeta=\zeta_{0}+\zeta_{1}\frac{\dot{a}}{a}+\zeta_{2}\frac{\ddot{a}}{\dot{a}}
\end{equation}
The first term is a constant $\zeta_{0}$, the second term is proportional to the Hubble
parameter, which characterizes the dependence of the bulk viscosity on velocity, and the third is proportional to $\frac{\ddot{a}}{\dot{a}}$, characterizing the effect of acceleration on the bulk viscosity.In terms of Hubble parameter
$H=\frac{\dot{a}}{a}$, this can be written as,
\begin{equation}
\label{z} \zeta=\zeta_{0}+\zeta_{1}H+\zeta_{2}(\frac{\dot{H}}{H}+H)
\end{equation}

The Friedmann equations governing the bulk viscous universe with cosmological constant are given as,
\begin{equation}
\label{friedmann1} H^{2}=\frac{\rho_{m}+\rho_\Lambda}{3}
\end{equation}
\begin{equation}
\label{friedmann2}
2\frac{\ddot{a}}{a}+\left(\frac{\dot{a}}{a}\right)^{2}=\rho_\Lambda-P^{*}
\end{equation}
where we have taken $8\pi G = 1$, $\rho_{m}$  and $\rho_\Lambda=\frac{\Lambda}{8\pi G}$ are the densities of matter and cosmological constant $\Lambda$, respectively and overdot represents the derivative with respect to cosmic time $t$. We consider separate conservation equations for matter and dark energy and are given below,
\begin{equation}
\label{conserm} 
\dot{\rho}_{m}+3H(\rho_{m}+P^{*})=0.
\end{equation}
\begin{equation}
\label{conserL} 
\dot{\rho}_{\Lambda}=0
\end{equation}
where we have assumed a constant equation of state for $\Lambda$, $\omega_\Lambda=-1$. Using the Friedmann equations (\ref{friedmann1}) and (\ref{friedmann2}) and using equations. (\ref{z}) and (\ref{p}), we get the differential equation for the Hubble parameter as,
\begin{equation}
\label{dH}
\dot{H}=\frac{1}{2-\tilde{\zeta}_2}\left(\tilde{\zeta}_0 HH_0+(\tilde{\zeta}_1+\tilde{\zeta}_2-3)H^2+3H_0^2 \Omega_{\Lambda 0}\right)
\end{equation}
where we have defined the dimensionless bulk viscous parameters $\tilde{\zeta}_0, \tilde{\zeta}_1, \tilde{\zeta}_2$ as,
\begin{equation}
\label{dimen} \tilde{\zeta}_{0}=\frac{3\zeta_{0}}{H_{0}},\ \ \ \
\tilde{\zeta}_{1}=3\zeta_{1},\ \ \ \ \tilde{\zeta}_{2}=3\zeta_{2}
\end{equation}
$H_{0}$ is the
present value of the Hubble parameter and $\Omega_{\Lambda 0}$ is the present density parameter of dark energy. Integrating equation (\ref{dH}) we can get the expression for the Hubble parameter as,
\begin{equation}
\label{H}
H=H_0\left[\frac{\left(y+\tilde{\zeta}_0\right)\left(y-2(\tilde{\zeta}_1+\tilde{\zeta}_2-3)-\tilde{\zeta}_0\right)e^{\frac{H_0(t-t_0)y}{2-\tilde{\zeta}_2}}-\left(y-\tilde{\zeta}_0\right)\left(y+2(\tilde{\zeta}_1+\tilde{\zeta}_2-3)+\tilde{\zeta}_0\right)}{2\left(\tilde{\zeta}_1+\tilde{\zeta}_2-3\right)\left(e^{\frac{H_0(t-t_0)y}{2-\tilde{\zeta}_2}}\left(-y+2(\tilde{\zeta}_1+\tilde{\zeta}_2-3)+\tilde{\zeta}_0\right)-\left(y+2(\tilde{\zeta}_1+\tilde{\zeta}_2-3)+\tilde{\zeta}_0\right)\right)}\right]
\end{equation}
where $y=\sqrt{\tilde{\zeta}_0^2-12\Omega_{\Lambda 0}(\tilde{\zeta}_1+\tilde{\zeta}_2-3)}$ and $t_0$ is the present cosmic time. As $t-t_0\rightarrow \infty$, $H\rightarrow H_0\left[\frac{y+\tilde{\zeta}_0}{2(\tilde{\zeta}_1+\tilde{\zeta}_2-3)}\right]$, a constant provided $\tilde{\zeta}_2<2$. When $t-t_0$ is small, H evolves as $H_0\left[\frac{2(2-\tilde{\zeta}_2)+H_0(t-t_0)(\tilde{\zeta}_0+6\Omega_{\Lambda 0}+y)}{2(2-\tilde{\zeta}_2)+H_0(t-t_0)(y-2(\tilde{\zeta}_1+\tilde{\zeta}_2-3)-\tilde{\zeta}_0)}\right]$. 

Using the definition of the Hubble parameter, we could obtain the expression for the scale factor from equation (\ref{H}) as,
\begin{equation}
\label{scalefactor}
a=e^{\frac{H_0(t-t_0)(y-\tilde{\zeta}_0)}{2(\tilde{\zeta}_1+\tilde{\zeta}_2-3)}}\left[\frac{y+2(\tilde{\zeta}_1+\tilde{\zeta}_2-3)+\tilde{\zeta}_0+e^{\frac{H_0(t-t_0)y}{2-\tilde{\zeta}_2}}(y-2(\tilde{\zeta}_1+\tilde{\zeta}_2-3)-\tilde{\zeta}_0)}{2y}\right]^{\frac{\tilde{\zeta}_2-2}{\tilde{\zeta}_1+\tilde{\zeta}_2-3}}
\end{equation}

When $\Omega_{\Lambda 0}=0$, the scale factor reduces to
\begin{equation}
a(t)=\left[(\frac{\tilde{\zeta}_{0}+\tilde{\zeta}_{12}-3}{\tilde{\zeta}_{0}})+(\frac{3-\tilde{\zeta}_{12}}{\tilde{\zeta}_{0}})
e^{\frac{\tilde{\zeta}_{0}}{2-\tilde{\zeta}_{2}}H_{0}(t-t_{0})}\right]^{\frac{2-\tilde{\zeta}_{2}}{3-\tilde{\zeta}_{12}}}
\end{equation}
which is the expression obtained in \cite{Athira1}. When $t-t_0$ is small, the scale factor evolves as
\begin{equation}
a\sim\left[1+\frac{H_0(t-t_0)(y-\tilde{\zeta}_0)}{2(\tilde{\zeta}_1+\tilde{\zeta}_2-3)}\right]\left[1+\frac{H_0(t-t_0)}{2-\tilde{\zeta}_2}(y-2(\tilde{\zeta}_1+\tilde{\zeta}_2-3)-\tilde{\zeta}_0)\right]^{\frac{\tilde{\zeta}_2-2}{\tilde{\zeta}_1+\tilde{\zeta}_2-3}}
\end{equation}
When $t-t_0$ is very large, from the expression of scale factor we see that it will increases exponentially.

\subsection{Equation of state and Deceleration parameter}
The equation of state parameter $\omega$ and the deceleration parameter $q$ can be obtained using the following relation,
\begin{equation}
\label{equation} \omega=-1-\frac{2}{3}\frac{\dot{H}}{H^2}
\end{equation}
\begin{equation}
\label{deceleration} q=-1-\frac{\dot{H}}{H^{2}}
\end{equation}
Using the expression (\ref{dH}) and (\ref{H}), we get the expressions for $\omega$ and $q$ as,
\begin{equation}
\label{equation_of_state}
\omega=-1+\frac{2y^2 \left(\tilde{\zeta}_0+\tilde{\zeta}_1+\tilde{\zeta}_2-3+3 \Omega _{\Lambda 0}\right) }{3 \left(\tilde{\zeta}_2-2\right) \left(\text{Sinh}\left[\frac{H_0 \left(t-t_0\right) y}{2 \left(2-\tilde{\zeta}_2\right)}\right] \left(\tilde{\zeta}_0+6 \Omega _{\Lambda 0}\right)+\text{Cosh}\left[\frac{H_0 \left(t-t_0\right) y}{2 \left(2-\tilde{\zeta}_2\right)}\right] y\right){}^2}
\end{equation}
\begin{equation}
\label{deceleration_parameter}
q=-1+\frac{y^2\left(\tilde{\zeta}_0+\tilde{\zeta}_1+\tilde{\zeta}_2-3+3 \Omega _{\Lambda 0}\right)}{\left(\tilde{\zeta}_2-2\right) \left(\text{Sinh}\left[\frac{H_0 \left(t-t_0\right) y}{2 \left(2-\tilde{\zeta}_2\right)}\right] \left(\tilde{\zeta}_0+6 \Omega _{\Lambda 0}\right)+\text{Cosh}\left[\frac{H_0 \left(t-t_0\right) y}{2 \left(2-\tilde{\zeta}_2\right)}\right] y\right){}^2}
\end{equation}

The present value of $\omega$ and $q$ can be obtained by putting $t=t_0$ and are,
\begin{equation}
\omega_0=\frac{2\tilde{\zeta}_0+2\tilde{\zeta}_1-\tilde{\zeta}_2+6\Omega_{\Lambda 0}}{3(\tilde{\zeta}_2-2)}
\end{equation}
\begin{equation}
q_0=\frac{\tilde{\zeta}_0+\tilde{\zeta}_1-1+3\Omega_{\Lambda 0}}{\tilde{\zeta}_2-2}
\end{equation}

The present universe will be accelerating only if $3\omega_0+1<0$ and $q_0<0$ and for the universe to be in quintessence region and to avoid big rip, it should satisfy the relation $q_0>-1$. Using these conditions and from the behaviour of the Hubble parameter and the scale factor, for a universe to begin from the big bang and then entering it to decelerated epoch and then making a transition to the accelerated epoch in the past, a set of conditions has to be satisfied by the $\tilde{\zeta}$'s. These conditions are,
\begin{enumerate}
\item $\tilde{\zeta}_0>0$,\, \, $\tilde{\zeta}_2<2$,\, \, $\tilde{\zeta}_0+\tilde{\zeta}_1>1-3\Omega_{\Lambda 0}$,\, \, $\tilde{\zeta}_1+\tilde{\zeta}_2<3$,\, \, $\tilde{\zeta}_0+\tilde{\zeta}_1+\tilde{\zeta}_2<3-3\Omega_{\Lambda 0}$

\item $\tilde{\zeta}_0<0$,\, \, $\tilde{\zeta}_2>2$,\, \, $\tilde{\zeta}_0+\tilde{\zeta}_1<1-3\Omega_{\Lambda 0}$,\, \, $\tilde{\zeta}_1+\tilde{\zeta}_2>3$,\, \, $\tilde{\zeta}_0+\tilde{\zeta}_1+\tilde{\zeta}_2>3-3\Omega_{\Lambda 0}$
\end{enumerate} 

If we neglect the cosmological constant i.e., $\Omega_{\Lambda 0}=0$, then these would reduce to the conditions obtained in the reference \cite{Athira1}. 

\section{With constant bulk viscosity}
\label{sec 2}
Let us consider the case when bulk viscous coefficient is a constant, i.e., when $\zeta=\zeta_0$ . The expression for Hubble parameter becomes,
\begin{equation}
H=H_0\frac{y-\tilde{\zeta }_0-6 \Omega _{\text{$\Lambda $0}}+e^{\frac{1}{2} H_0 \left(t-t_0\right) y} \left(y+\tilde{\zeta} _0+6 \Omega _{\text{$\Lambda $0}}\right)}{y+\tilde{\zeta} _0-6+e^{\frac{1}{2} H_0 \left(t-t_0\right) y} \left(y-\tilde{\zeta} _0+6\right)}
\end{equation}
where $y=\sqrt{\tilde{\zeta}_0^2+36\Omega_{\Lambda 0}}$
Similarly, one could obtained the expression for scale factor for constant $\zeta$ as, 
\begin{equation}
\label{scalefactor_Lambda0}
a=e^{\frac{1}{6} H_0\left(t-t_0\right) \left(\tilde{\zeta} _0-y\right)}\left(\frac{\left(y+\tilde{\zeta} _0-6\right)+e^{\frac{H_0\left(t-t_0\right)y}{2}} \left( y-\tilde{\zeta} _0+6\right)}{2 y}\right)^{\frac{2}{3}}
\end{equation}
Similarly, the corresponding equation of state and the deceleration parameter for constant viscosity becomes,
\begin{equation}
\omega=\left(-1-\frac{ \left(\tilde{\zeta} _0-3+3 \Omega _{\Lambda 0}\right) y^2}{3\left(y \text{Cosh}\left[\frac{1}{4} H_0 \left(t-t_0\right) y\right]+\left(\tilde{\zeta} _0+6 \Omega _{\Lambda 0}\right) \text{Sinh}\left[\frac{1}{4} H_0 \left(t-t_0\right) y\right]\right){}^2}\right)
\end{equation}
\begin{equation}
q=\left(-1-\frac{\left(\tilde{\zeta} _0-3+3 \Omega _{\Lambda 0}\right) y^2}{2\left(y \text{Cosh}\left[\frac{1}{4} H_0 \left(t-t_0\right)y\right]+\left(\tilde{\zeta} _0+6 \Omega _{\Lambda 0}\right) \text{Sinh}\left[\frac{1}{4}H_0 \left(t-t_0\right)y\right]\right){}^2}\right)
\end{equation}

As mentioned before, for an accelerating universe, the present value of equation of state $\omega_0<\frac{-1}{3}$ and the present value of the deceleration parameter $q_0<0$. To avoid big rip, the equation of state parameter $\omega_0>-1$, above the phantom limit. These conditions help us to constrain the value of $\tilde{\zeta}_0$ as 
\begin{equation}
\label{constrain_with_Lambda0}
1-3\Omega_{\Lambda 0}<\tilde{\zeta}_0<3(1-\Omega_{\Lambda 0}).
\end{equation}

From observation $\Omega_{\Lambda}$ is constrained in the range $0.65-0.75$ \cite{}. This constrains the $\tilde{\zeta}_0$ in between $-1.25<\tilde{\zeta}_0<1.05$.
\subsection{Age of the universe}
\begin{figure}[tbp]
\centering
\includegraphics[width=.6\textwidth]{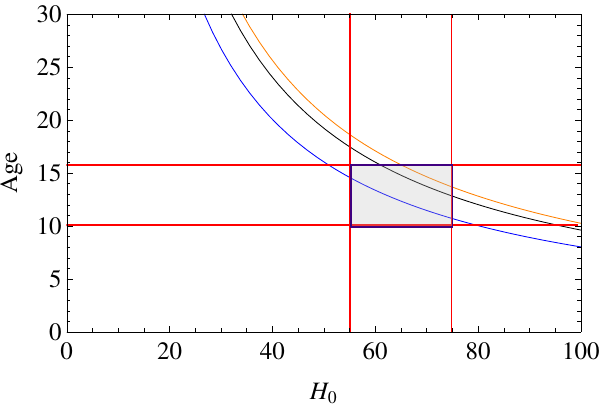}
\caption{\label{age_Lambda0} The figure shows the variation of age with $H_0$ for different values of 
$(\tilde{\zeta}_0,\Omega_{\Lambda})$. Black line corresponds to $(\tilde{\zeta}_0,\Omega_{\Lambda})=(0.1,0.68)$. 
The orange line and blue line corresponds to $(\tilde{\zeta}_0,\Omega_{\Lambda})=(0.2,0.7)$ and $(-0.5,0.7)$ respectively.}
\end{figure}

Age of the universe in this case can be obtained by equating $a=1$ in the equation \eqref{scalefactor_Lambda0} and is 
found to be,
\begin{equation}
\textrm{Age}\equiv \left(\frac{2}{H_0 y}\right)\text{Log}\left[1-\frac{2 y}{6+y-\tilde{\zeta}_0}\right].
\end{equation} 
The plot of age of the universe for different values of $(\tilde{\zeta}_0,\Omega_{\Lambda})$ subjected to
the constrain (\ref{constrain_with_Lambda0}) are shown in the figure \eqref{age_Lambda0}.
The age plot shows reasonably good agreement for $(\zeta_0,\Omega_{\Lambda})=(-0.5,0.7)$ but the agreement with respect 
$(\zeta_0,\Omega_{\Lambda})=(0.1,0.68)$ is slightly less and for the third choice it is not in nice agreement.
But corresponding to the best agreement pair the viscosity is negative. Whether is physically feasible or not may evident from 
the further considerations of the entropy evolution and dynamical system behaviour.

\subsection{Thermodynamics}
\begin{figure}
\centering
\includegraphics[scale=0.8]{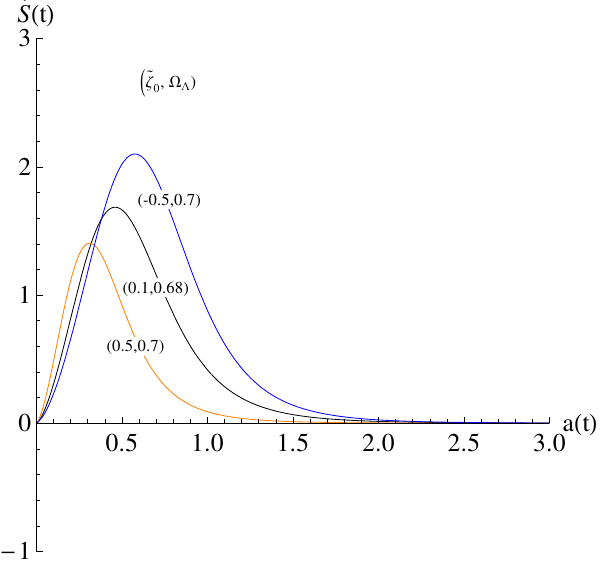}
\caption{\label{fig:first_entropy_Lambda} Evolution of the first derivative of entropy with the scale factor for different values of $(\tilde{\zeta}_0,\Omega_{\Lambda})$ subjected to the constrain (\ref{constrain_with_Lambda0}).}
\end{figure}

\begin{figure}
\centering
\includegraphics[scale=0.8]{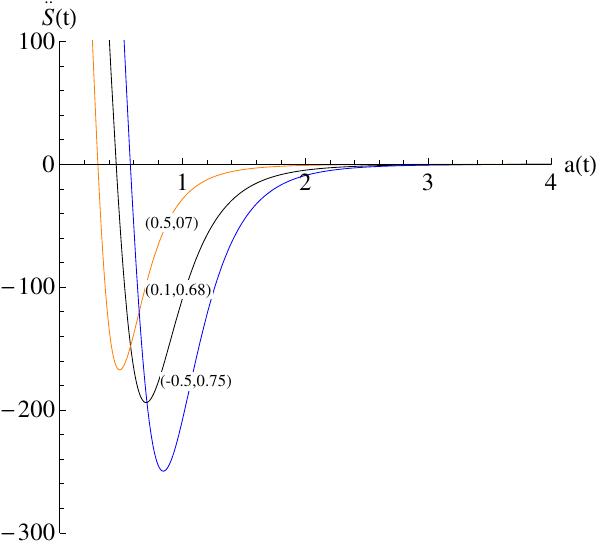}
\caption{\label{fig:second_entropy_Lambda} Evolution of the second derivative of entropy with the scale factor for different values of $(\tilde{\zeta}_0,\Omega_{\Lambda})$ subjected to the constrain (\ref{constrain_with_Lambda0}).}
\end{figure}

We now check the validity of the Generalized second law and maximization of entropy condition in this case. 
Assuming apparent horizon as the boundary of the universe and obtaining the horizon entropy using the 
Bekenstein relation  and matter entropy using the Gibbs equation, we calculated the expression for the first derivative and second derivative of the total 
entropy with respect to time. The relation obtained are as follows:
\begin{equation}
\dot{S}=\frac{64 \pi ^2e^{t' \tilde{y}}b^2 \tilde{y}^4 \left(\tilde{y}-6+\tilde{\zeta}_0 +e^{\frac{1}{2} t' \tilde{y}} (\tilde{y}+6-\tilde{\zeta}_0)\right)}{H_0 \left(\tilde{y}-\tilde{\zeta}_0 -6 \Omega_{\Lambda} +e^{\frac{1}{2} t' \tilde{y}} (6 \Omega_{\Lambda} +\tilde{y}+\tilde{\zeta}_0)\right)^5},
\end{equation}
\begin{equation}
\ddot{S}=-\frac{384 \pi ^2 b^2 \tilde{y}^5 e^{\frac{3}{2}t' \tilde{y}}(b \tilde{y}+2 (1+\Omega_{\Lambda}) \tilde{y} \text{Cosh}[\frac{1}{2}t' \tilde{y}]+2 d \text{Sinh}[\frac{1}{2}t'\tilde{y}])}{((-1+e^{\frac{1}{2}t'\tilde{y}})\tilde{\zeta}_0 -6 \Omega_{\Lambda}+\tilde{y}+e^{\frac{1}{2}t'\tilde{y}} (6 \Omega_{\Lambda} +\tilde{y}))^6},
\end{equation}
where $b=\tilde{\zeta}_0+3\Omega_{\Lambda 0}-3$, $d=\tilde{\zeta}_0+12 \Omega_{\Lambda}-\tilde{\zeta}_0\Omega_{\Lambda}$ and $t'=H_0(t-t_0)$. 
The evolution of $\dot{S}$ and $\ddot{S}$ with respect to the scale factor for different values of $\Omega_{\Lambda}$ and $\tilde{\zeta}_0$ subjected to the constrain \eqref{constrain_with_Lambda0} are plotted and are shown in figures (\ref{fig:first_entropy_Lambda}) and (\ref{fig:second_entropy_Lambda}) respectively
From the figures, it is clear that GSL and maximization of entropy condition is valid for the model.

\subsection{Phase space analysis}
We also try to study the asymptotic behavior of the model. We chose $u$ and $v$ as the phase space variables defined as
\begin{equation}
\begin{split}
\begin{aligned}
u&=\Omega_{m}=\frac{\rho_{m}}{3H^{2}},\\
v&=\frac{1}{\frac{H_{0}}{H}+1},
\end{aligned}
\end{split}
\end{equation}
which varies in the range $0\leq
u\leq1$ and $0\leq v\leq1$.
Using the conservation equation and differential equation for Hubble parameter, we can obtained the autonomous equations for $u$ and $v$ as,
\begin{equation}
\begin{aligned}
\begin{split}
u'&=\frac{(1-v)}{v^2}(v(1-u)\tilde{\zeta}_0 -3\Omega_{\Lambda} u (1-v)),\\
v'&=\frac{(1-v)}{2 v}\left(3\Omega_{\Lambda}(1-v)^2+\tilde{\zeta}_0 v(1-v)-3v^2\right).
\end{split}
\end{aligned}
\end{equation}
There are three critical points for the above autonomous equation and the corresponding eigen values are listed in the Table \ref{tab:1Lambda}.
\begin{table}[tbp]
\centering
\begin{tabular}{|c|c|}
\hline
$\left(u_c,v_c\right)$ & Eigen value $\left(\lambda_1,\lambda_2\right)$ \\
\hline
$\left(u,1\right)$ & $\left(\frac{3}{2},0\right)$ \\
$\left(-\frac{\tilde{\zeta}_0\left(\tilde{\zeta}_0 +\sqrt{\tilde{\zeta}_0 ^2+36 \Omega_{\Lambda} }\right)}{18\Omega_{\Lambda} },\frac{\tilde{\zeta}_0 -6 \Omega_{\Lambda} -\sqrt{\tilde{\zeta}_0 ^2+36 \Omega_{\Lambda}}}{6+2 \tilde{\zeta}_0 -6 \Omega_{\Lambda} }\right)$ & $\left(-3,\frac{3}{-1+\frac{\tilde{\zeta}_0 }{\sqrt{\tilde{\zeta}_0^2+36\Omega_{\Lambda} }}}\right)$\\
$\left(\frac{\tilde{\zeta}_0   \left(-\tilde{\zeta}_0 +\sqrt{\tilde{\zeta}_0  ^2+36 \Omega_{\Lambda} }\right)}{18 \Omega_{\Lambda} }, \frac{\tilde{\zeta}_0  -6 \Omega_{\Lambda} +\sqrt{\tilde{\zeta}_0  ^2+36 \Omega_{\Lambda} }}{6+2 \tilde{\zeta}_0  -6\Omega_{\Lambda} }\right)$ & $\left(-3,\frac{3}{-1-\frac{\tilde{\zeta}_0 }{\sqrt{\tilde{\zeta}_0 ^2+36\Omega_{\Lambda} }}}\right) $\\
\hline
\end{tabular} 
\caption{\label{tab:1Lambda} Critical values and the corresponding eigen values for the bulk viscous model with $\Lambda$ for $\zeta=\zeta_0$}
\end{table}
Inorder to represent a universe with unstable matter dominated phase and a stable, physically feasible accelerated 
phase we see that $\tilde{\zeta}_0$ must be positive subjected to the constrain \eqref{constrain_with_Lambda0}. 
In determining the age corresponding to this model we have noted that, the best fit have arised both with negative value 
of $\zeta_0$ and also with positive value (the black line in the age plot) of $\zeta_0.$ But the asymptotic analysis presented 
here, however supports only a positive value for $\zeta_0.$ Earlier in the analysis without cosmological constant also we conclude 
that, the case with $\zeta=\zeta_0$ is preferred over other cases. Thus even though the age 
prediction has been changed slightly, the present model is also predicting a conventional evolution of the universe with 
constant viscosity as in the case of the model without cosmological constant.

\section{With $\zeta=\zeta_1H$}
\label{sec 3}
Let us consider another special case of $\zeta=\zeta_1 H$. So here $\zeta$ depends only on the velocity component of the expansion of the universe. The expression for the Hubble Parameter and the scale factor are as follows,
\begin{equation}
H=-\frac{\sqrt{3} H_0\Omega_{\Lambda 0} \left(6-2 \tilde{\zeta} _1-2\sqrt{3(3-\tilde{\zeta} _1)\Omega_{\Lambda 0}}+2e^{ H_0(t-t_0) \sqrt{3(3-\tilde{\zeta} _1)\Omega_{\Lambda 0}}}(3-\tilde{\zeta} _1+\sqrt{3(3-\tilde{\zeta} _1)\Omega_{\Lambda 0}})\right)}{\sqrt{(3-\tilde{\zeta} _1)\Omega_{\Lambda 0}} \left(6-2 \tilde{\zeta} _1-2\sqrt{3(3-\tilde{\zeta} _1)\Omega_{\Lambda 0}}-2 e^{ H_0(t-t_0) \sqrt{3(3-\tilde{\zeta }_1)\Omega_{\Lambda 0}}} (3-\tilde{\zeta} _1+ \sqrt{3(3-\tilde{\zeta} _1)\Omega_{\Lambda 0}})\right)}
\end{equation}
\begin{equation}
\l
a=12^{\frac{1}{\tilde{\zeta}_1-3}}e^{-\frac{\sqrt{3} H_0 (t-t_0)\Omega_{\Lambda 0}}{\sqrt{(3-\tilde{\zeta}_1)\Omega_{\Lambda 0}}}} \left(\frac{\tilde{\zeta}_1-3+\sqrt{3(3-\tilde{\zeta}_1)\Omega_{\Lambda 0}}+e^{ H_0(t-t_0)\sqrt{3(3-\tilde{\zeta}_1)\Omega_{\Lambda 0}}}\left(3-\tilde{\zeta}_1+\sqrt{3(3-\tilde{\zeta}_1)\Omega_{\Lambda 0}}\right)}{\sqrt{(3-\tilde{\zeta}_1)\Omega_{\Lambda 0}}}\right)^{\frac{2}{3-\tilde{\zeta}_1}}
\end{equation} 
From the expression of Hubble parameter and the scale factor, we see that inorder to represent the conventional behavior of the universe, $\tilde{\zeta}_1$ should be less than 3. In this case one could obtain the expression for the Hubble parameter in terms of the scale factor $a$. And it is found to be,
\begin{equation}
\label{H(a)}
H=H_0\sqrt{\left[\frac{a^{\tilde{\zeta}_1-3}(\tilde{\zeta}_1-3+3\Omega_{\Lambda 0})-3\Omega_{\Lambda 0}}{\tilde{\zeta}_1-3}\right]}
\end{equation}
Since a direct relation between the Hubble parameter $H$ and the scale factor $a$ is found out, it is possible to extract the value of $\zeta_1$.
\subsection{Extraction of $\tilde{\zeta}_1$}
To extract the value of $\tilde{\zeta}_1$, we use the latest Pantheon Type Ia Supernova data consisting of 1048 data points.. The method used is the $\chi^2$ minimization technique and is defined as,
\begin{equation}
\label{chisquare}
\chi^{2}\equiv
\sum^{n}_{k=1}\frac{\left[\mu_{t}-\mu_{k}\right]^{2}}{\sigma_{k}^{2}},
\end{equation}
where $\mu_{k}$ is the observational distance modulus for the k-th
Supernova (obtained from the data) with red shift $z_k$, $\sigma_{k}^{2}$ is the variance of the measurement,
$n$ is the total number of data and $\mu_{t}$ is the theoretical distance modulus for the
k-th Supernova with the same redshift $z_{k}$, which is given as
\begin{equation}
\mu_{t}=m-M=5\log_{10}[\frac{d_{L}}{Mpc}]+25
\end{equation}
where, $m$ and $M$ are the apparent and absolute magnitudes of the
SNe respectively. $d_{L}$ is the luminosity distance and is defined as
\begin{equation}
d_{L}=c(1+z)\int_{0}^{z}\frac{dz'}{H},
\end{equation}
where $c$ is the speed of light. Using the expression for $H$ from equation (\ref{H(a)}), we construct the $\chi^2$ function. 
We extract the values of $\Omega_{\Lambda 0}$ and $H_0$ along with $\tilde{\zeta}_1$. The values are given in the table below \ref{tab:1}.
\begin{table}[tbp]
\centering
\begin{tabular}{|l|l|l|l|l|l|l|}
\hline\noalign{\smallskip}
Model & $\tilde{\zeta}_1$ & $H_0$ & $\Omega_{\Lambda 0}$ & $\Omega_{m0}$ & $\chi^2_{min}$ & $\chi^2_{d.o.f}$ \\
\noalign{\smallskip}\hline\noalign{\smallskip}
$\zeta=\zeta_{1}\frac{\dot{a}}{a}$ & 0.351 & 69.5 & 0.75 & 0.25 & 1033 & 0.9 \\
\noalign{\smallskip}\hline
\end{tabular} 
\caption{\label{tab:1} Best estimates of the bulk viscous parameter $\tilde{\zeta}_1$, $H_{0}$, $\Omega_{\Lambda 0}$, $\Omega_{m0}=1-\Omega_{\Lambda 0}$ and also $\chi^{2}$ minimum value for $\zeta=\zeta_{1}\frac{\dot{a}}{a}$
$\chi^{2}_{d.o.f}=\frac{\chi^{2}_{min}}{n-m}$, where $n=307$, the
number of data and $m$ is the number of parameters in the model. The
subscript d.o.f stands for degrees of freedom. For the best
estimation we have used Pantheon supernova data set consisting of 1048 data points.}
\end{table}
\subsection{Evolution of equation of state parameter and deceleration parameter}
\begin{figure}
\centering
\includegraphics[scale=0.8]{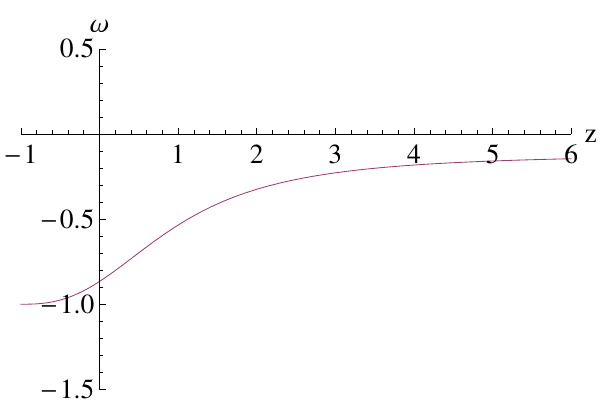}
\caption{\label{fig:w_z1_L} Plot of the equation of state with the redshift for the best estimated values of  $\tilde{\zeta}_1$ and $\Omega_{\Lambda}$.}
\end{figure}

\begin{figure}
\centering
\includegraphics[scale=0.8]{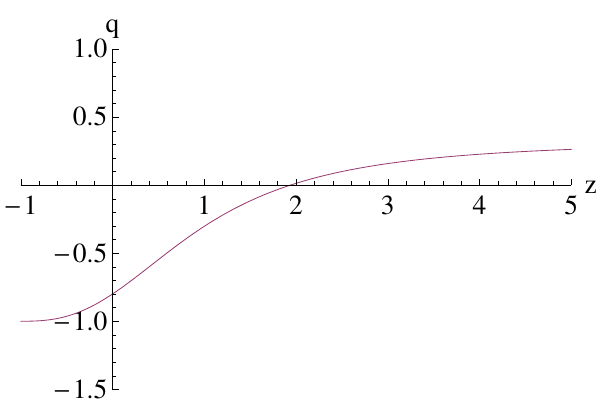}
\caption{\label{fig:q_z1_L} Plot of the deceleration parameter with the redshift for the best estimated values of  $\tilde{\zeta}_1$ and $\Omega_{\Lambda}$.}
\end{figure}

The expression for the equation of state parameter and the deceleration parameter for this model can be obtained by making $\tilde{\zeta}_0=\tilde{\zeta}_2=0$ in the equations (\ref{equation_of_state}) and  (\ref{deceleration_parameter}) respectively.
\begin{equation}
\omega=-1-\frac{(\tilde{\zeta}_1-3)(\tilde{\zeta}_1-3+3 \Omega_{\Lambda 0})}{\left(\sqrt{3(\tilde{\zeta}_1-3)} \text{Cos}[\frac{1}{2} H_0 (t-t_0) \sqrt{3\Omega_{\Lambda 0}(\tilde{\zeta}_1-3)}] +3\sqrt{\Omega_{\Lambda 0}} \text{Sin}[\frac{1}{2}H_0 (t-t_0) \sqrt{3\Omega_{\Lambda 0}(\tilde{\zeta}_1-3)}] \right)^2}
\end{equation}
\begin{equation}
q=-1-\frac{3 (\tilde{\zeta}_1-3)(\tilde{\zeta}_1-3+3 \Omega _{\Lambda 0})}{2 \left(\sqrt{3(\tilde{\zeta}_1-3)} \text{Cos}[\frac{1}{2}H_0(t-t_0)\sqrt{3\Omega _{\Lambda 0}(\tilde{\zeta}_1-3)}] +3 \sqrt{\Omega_{\Lambda 0}}\text{Sin}[\frac{1}{2} \sqrt{3} H_0 (t-t_0) \sqrt{\tilde{\zeta}_1-3} \sqrt{\Omega_{\Lambda 0}}] \right)^2}
\end{equation}
The equation of state parameter $\omega$ and the deceleration parameter $q,$ in terms of scale factor  are given as,
\begin{equation}
\label{eqn:w_z1_L}
\omega=\frac{9 a^3 \Omega_{\Lambda 0} -a^{\tilde{\zeta}_1} \tilde{\zeta}_1  (\tilde{\zeta}_1 -3+3 \Omega_{\Lambda 0})}{-9 a^3 \Omega_{\Lambda 0} +3 a^{\tilde{\zeta}_1} (\tilde{\zeta}_1 -3+3 \Omega_{\Lambda 0})},
\end{equation}
\begin{equation}
\label{eqn:q_z1_L}
q=-1-\frac{a^{\tilde{\zeta}_1} (-3+\tilde{\zeta}_1) (\tilde{\zeta}_1 -3+3 \Omega_{\Lambda} )}{-6 a^3 \Omega_{\Lambda} +2 a^{\tilde{\zeta}_1} (\tilde{\zeta}_1-3+3 \Omega_{\Lambda} )}.
\end{equation}
The plot of $\omega$ and $q$ for the best estimated values of $\tilde{\zeta}_1$ and $\Omega_{\Lambda}$ are shown 
in the figures \ref{fig:w_z1_L} and \ref{fig:q_z1_L} respectively.
The equation of state is zero in the recent past. It decreases to the 
negative values and finally saturated at $\omega=-1$ corresponding to a de Sitter epoch in the extreme future. The evolution of the deceleration parameter starts from around $q \sim 0.5$ in the past, which corresponds 
to decelerated epoch and decreasing as the universe expands. It saturates at $q=-1$ corresponding the future de Sitter phase.
The present value of $\omega$ and $q$ can be obtained by putting $a=1$ in the expressions given by equation \eqref{eqn:w_z1_L} and \eqref{eqn:q_z1_L}, respectively and are obtained as,
\begin{equation}
\omega_0=-\frac{\tilde{\zeta}_1 }{3}-\Omega_{\Lambda},
\end{equation}
\begin{equation}
q_0=\frac{1}{2} (1-\tilde{\zeta}_1 -3\Omega_{\Lambda} ).
\end{equation} 
Using the best estimated values of $\tilde{\zeta}_1$ and $\Omega_{\Lambda}$, we get $\omega_0=-0.867033$ and $q_0=-0.80055$, which is near to concordance value obtained by WMAP observation.

\subsection{Age of the universe}
The age of the universe in this model can be obtained by equating the scale factor (equation \eqref{H(a)}) to one and is found to be
\begin{equation}
Age\equiv\frac{\text{Log}\left[\frac{3-\tilde{\zeta}_1-\sqrt{3} \sqrt{\left(3-\tilde{\zeta}_1\right) \Omega _{\Lambda 0}}}{3-\tilde{\zeta}_1+\sqrt{3} \sqrt{\left(3-\tilde{\zeta}_1\right) \Omega _{\Lambda 0}}}\right]}{\sqrt{3} p \sqrt{-\left(-3+\tilde{\zeta}_1\right) \Omega _{\Lambda 0}}}.
\end{equation}
Using the best estimated values for  $\tilde{\zeta}_1$ and $\Omega_{\Lambda}$, the age is found to $18.44$Gyr and is matching with 
the concordance value of the age of the universe obtained from the oldest globular observations. In this way the model is promising 
in predicting the age. 
\section{Conclusion}
\label{sec 4}
We analyse a universe with a cosmological constant and bulk viscous matter. By considering the general form for $\zeta=\zeta_{0}+\zeta_{1}\frac{\dot{a}}{a}+\zeta_{2}\frac{\ddot{a}}{\dot{a}}$, we obtain the constrains of the viscous parameters by finding the evolution of Hubble parameter, scale factor and cosmological parameters. 

Two special cases for the viscous coefficient $\zeta$, $\zeta=\zeta_0$, a constant and $\zeta=\zeta_1 H$, depending on the velocity of the expanding universe are considered. For $\zeta=\zeta_0$, for the constrain is $-1.25<\tilde{\zeta}_0<1.05$. It is also found out that under this constrain the age of the universe is in accordance with the galactic observations. GSL and maximization of entropy condition are also found to be valid for the model.

For $\zeta=\zeta_1H$, the value on $\zeta_1$ is extracted using pantheon data and is found to be $0.351$. The present value of deceleration parameter and equation of state is found to be $q_0=-0.80055$ and $\omega_0=-0.867033$, respectively, which is near to concordance value obtained by WMAP observation. The age is found to $18.44$Gyr and is matching with the observations.

The addition of cosmological constant in the bulk viscous matter dominated universe improves age of the universe as well as other cosmological parameters.

\end{document}